\documentclass[twocolumn, times]{aastex701}
\usepackage{natbib}
\usepackage{placeins}
\usepackage{graphicx}
\usepackage{xcolor}
\usepackage{gensymb}
\usepackage{amsmath}
\usepackage{siunitx}
\usepackage{float}
\usepackage{bm}
\usepackage{xspace}
\usepackage{comment}

\submitjournal{ApJ}

\shorttitle{Binaries are Jerks}

\begin{document}

\title{{Identifying Compact Object Candidates using Gaia DR3 Astrometric Accelerations and Jerks}}

\author[0009-0004-7683-157X]{Andrew B. Kinnion}
\affiliation{Department of Physics, University of Florida, 2001 Museum Rd., Gainesville, FL 32611}
\email{andrewkinnion@ufl.edu}

\author[0000-0001-5261-3923]{Jeff J. Andrews} 
\email{jeffrey.andrews@ufl.edu}
\affiliation{Department of Physics, University of Florida, 2001 Museum Rd., Gainesville, FL 32611}
\affiliation{Institute for Fundamental Theory, 2001 Museum Rd, Gainesville, FL 32611, USA}

\author[0009-0009-1888-8785]{Eirini Kasdagli}
\affiliation{Department of Physics, University of Florida, 2001 Museum Rd., Gainesville, FL 32611}
\email{kasdaglie@ufl.edu}

\begin{abstract}
Gaia Data Release 3 (DR3) provides astrometric measurements for more than a billion sources, hundreds of thousands of which have non-single-star astrometric solutions that indicate deviations from linear proper motion. Although sources with orbital solutions have been extensively inspected for compact objects via the astrometric mass function, two other solution types---the 7-parameter constant-acceleration solutions characterized by $\dot{\bm{\mu}}$, and the 9-parameter variable-acceleration solutions characterized by $\dot{\bm{\mu}}$ and $\ddot{\bm{\mu}}$---have not been as carefully explored. In this work, we develop a methodology for selecting sources within the Gaia DR3 9-parameter non-single-star astrometric solutions that are most likely to host massive compact companions.
We assume that a source is the luminous star in a binary orbit around a dark object and derive analytic expressions for the sky-projected accelerations and jerks for three binary populations with different orbit approximations: inclined eccentric, face-on eccentric, and face-on circular. Using \texttt{POSYDON}, we develop a fourth, more realistic population of binaries, each consisting of a luminous star and a black hole. For each population, we address observational biases using \texttt{gaiamock} and select only the systems it fits with a 9-parameter solution. We find that binaries likely to host massive compact companions can be identified from the characteristic joint astrometric accelerations and jerks exhibited by the luminous stars, despite 9-parameter fits under-constraining full orbits. Finally, to illustrate the utility of our approach, we analyze a historical object of interest, HD~12871, and examine 9-parameter fits of Gaia~BH3-like binaries.
\end{abstract}


\section{Introduction}
\label{sec:introduction}
Astrometry reveals the dynamics and structure of the Milky Way by providing measurements of the positions and proper motions of its stellar constituents. The first satellite to focus on astrometry was Hipparcos, which 
measured positions with milliarcsecond-level precision~\citep[see][]{1984MitAG..62...63K}. Its largest releases were the Hipparcos Catalog~\citep{1997A&A...323L..49P} and the Tycho-2 Catalog~\citep{2000A&A...355L..27H}, together containing more than two million stars. However, both catalogs are dwarfed by the over 100~billion stars estimated to exist in the Milky Way---leaving the vast majority of the Galactic population unprobed. Now, more than 30~years later, Gaia has ushered in a new era of astrometry with its microarcsecond-level precision~\citep{annurev:/content/journals/10.1146/annurev-astro-112320-035628}. Since its launch in 2013, there have been three data releases: Gaia DR1~\citep{2016A&A...595A...2G}, DR2~\citep{2018A&A...616A...1G}, and DR3~\citep{2023A&A...674A...1G}, with DR4 scheduled for the end of 2026 and DR5 for 2031. Each subsequent release contains an expanded catalog and improved precision due to a longer baseline of observations and thus an increased number of observations~\citep{2017sf2a.conf..259K}. To compare, DR3 contains more than 1.8~billion sources, while the upcoming DR4 is expected to contain 2.8~billion sources.

Gaia has made exciting discoveries for a wide range of subfields, including variable stars \citep{2023A&A...674A..13E}, Solar System objects \citep{2023A&A...674A..12T}, star clusters \citep{2018A&A...618A..93C}, and Milky Way dynamics \citep{2018A&A...616A..11G}. Furthermore, foundational binary system discoveries emerge from the sample of $\sim10^5$~astrometric binaries identified by fitting their observed orbital motion in the sky \citep{2023A&A...674A...9H}. In particular, a rare fraction of these fits suggest that they contain a compact object (CO) companion: a white dwarf (WD), neutron star (NS), or black hole (BH) \citep[for a review, see][]{2024NewAR..9801694E}. COs in binaries are typically detected through the accretion of matter from their companions or, in the case of pulsar binaries, through the Doppler motion of the pulsar's regular radio pulses. Identifying CO binaries through radial velocity (RV) variations has long been pursued \citep{1966ApJ...144..840Z, 1969ApJ...156.1013T, 1992ARA&A..30..287C, 2019Sci...366..637T}, but nearly every claimed CO binary detection using RVs alone was later shown to host a non-degenerate companion \citep[e.g.,][]{2020MNRAS.493L..22E, 2022MNRAS.515.1266E, 2022MNRAS.512.5620E}.

Astrometry, with full orbits, removes the notorious $\sin(i)$ degeneracy by measuring a star's motion in the plane of the sky and deriving Thiele-Innes elements, from which the inclination can be determined \citep{1975ARA&A..13..295V, 2002ApJ...572..944G, 2010ApJ...719..958T, 2014arXiv1407.6163B}. By confirming Gaia astrometric orbital solutions using RV follow-up observations, three BHs have been discovered in detached binary systems \citep{2023MNRAS.518.1057E, 2023MNRAS.521.4323E, 2024A&A...686L...2G}, more than two dozen NSs \citep{2024OJAp....7E..27E, 2024OJAp....7E..58E}, and thousands of WDs~\citep{2023ApJ...954....4G, 2024MNRAS.529.3729S, 2024MNRAS.529.4840G, 2024PASP..136h4202Y, 2024ApJ...964..101Z}. These COs provide key information on stellar and binary evolutionary parameters, including constraints on post-asymptotic giant branch mass transfer \citep{2025PASP..137j4205Y}, natal kicks \citep{2023MNRAS.518.1057E, 2024A&A...690A.144I, 2024MNRAS.535.3577K}, and common envelope efficiency \citep{2025A&A...704A...6R}.

Previous searches for COs within Gaia data have focused on single-line spectroscopic sources \citep{2023MNRAS.521.5927J}, metal-poor sources that exhibit RV variability \citep{2026ApJ..1000..148L}, sources exhibiting photometric variability \citep{2022ApJ...940..126F, 2023MNRAS.524.4367N, 2024ApJ...975..163C, 2024OJAp....7E..24R}, and sources with astrometric orbital solutions \citep{2022arXiv220700680A, 2023MNRAS.518.2991S, 2024MNRAS.52711719Y, 2024MNRAS.529.3729S}. However, very little attention has been paid to sources that have 7-parameter (involving $\dot{\bm{\mu}}$) or 9-parameter (involving both $\dot{\bm{\mu}}$ and $\ddot{\bm{\mu}}$) astrometric solutions. In one of the only studies to include a few such sources, \citet{Simon_2026} obtained follow-up observations of various candidate targets, of which 11 exhibited astrometric accelerations, finding only one that may host a neutron star. \citet{Simon_2026} also required these accelerating sources to have linear or quadratic RV fits, as well as large RV variations, which exclude a large portion of the full DR3 accelerating sources.

In this work, we examine sources within DR3 that exhibit both astrometric accelerations and astrometric jerks without using any additional parameter criteria. These sources are in the DR3 \texttt{nss\_acceleration\_astro} table having \texttt{nss\_solution\_type} ``Acceleration9'' (which we call Astro-9). Our goal is to determine metrics that will efficiently identify systems that are most likely to host compact object companions in order to optimize follow-up observational campaigns.
In Section~\ref{sec:methods}, we give an overview of orbital acceleration and jerk and formulate their projections onto the plane of the sky. In Section~\ref{sec:model}, we describe a rudimentary Milky Way and how we generate mock binary systems. 
We describe results from these mock populations, along with a population that accounts for binary evolution using the binary population synthesis code {\tt POSYDON} in Section~\ref{sec:results}. In Section~\ref{sec:discussion}, we describe how our results can be used to efficiently focus observational follow-up campaigns and include discussions of HD~12871, a historical CO candidate with a spectroscopic orbit from the literature, and simulations of Gaia~BH3-like systems. Finally, we provide our conclusions in Section~\ref{sec:Conclusion}.

\section{Analytic Methods}
\label{sec:methods}
We start by revisiting orbital dynamics to calculate orbital acceleration and jerk for a binary with arbitrary orbital period $P$, eccentricity $e$, luminous star mass $M_1$, and dark companion mass $M_2$. Then, we formulate the projection onto the plane of the sky, followed by the conversion to physical units. Finally, we provide explicit expressions that relate masses to orbital parameters for inclined and face-on orbital orientations.

\subsection{Orbital Acceleration and Jerk}
\label{sec:orbital_accel}
Newton's law of gravitation allows us to find the gravitational acceleration of one object towards another. For a bound orbit, we may refer to it as orbital acceleration, which we will simply call acceleration (and likewise jerk for the orbital jerk). Adopting the barycenter as the origin and using polar coordinates, the acceleration of the luminous star, $\ddot{\bm{r}}$, is
\begin{equation}
\label{eq:vector_accel}
\ddot{\bm{r}} = -\frac{GM_2}{r^2}\bm{\hat{r}},
\end{equation}
where $G$ is the gravitational constant, $r$ is the distance between the two objects, and $\hat{\bm{r}}$ is the unit position vector pointing from the barycenter to the luminous star. We define the azimuthal vector $\bm{f}=f\hat{\bm{f}}$ where $f$ is the true anomaly so that a point $(r, 0\degree)$ corresponds to a vector that points from the barycenter to the periastron of the luminous star. From the time-derivative of Equation~\ref{eq:vector_accel}, the jerk, $\dddot{\bm{r}}$, experienced by the luminous star is 
\begin{equation}
\dddot{\bm{r}} = \frac{GM_2}{r^2} \left(\frac{2\dot{r}}{r}\bm{\hat{r}}-\dot{f}\bm{\hat{f}} \right).
\end{equation}
It is compact to express the jerk of the luminous star using the magnitude of its acceleration so that
\begin{equation}
\label{eq:vector_jerk}
\frac{\dddot{\bm{r}}}{\lvert\ddot{\bm{r}}\rvert} = \frac{2\dot{r}}{r}\bm{\hat{r}}-\dot{f}\bm{\hat{f}}.
\end{equation}
The cosine of the angle between the acceleration and jerk vectors, $\theta$, is
\begin{equation}
\label{eq:true_aj_ang}
\cos \theta=\frac{-2\dot{r}}{\sqrt{4\dot{r}^2+r^2\dot{f}^2}}.
\end{equation}
For a given $f$, $e$, and semi-major axis $a$, the separation between the two objects is
\begin{equation}
\label{eq:r}
r = \frac{a\left(1-e^2\right)}{\left(1+e\cos f \right)}.
\end{equation}

In order to use a rotation matrix to obtain the projection onto the plane of the sky, the radial and angular unit vectors, $\bm{\hat{r}}$ and $\bm{\hat{f}}$, respectively, are converted into their Cartesian representations in the orbital plane with basis vectors $\hat{X}$ in the direction of periastron of the luminous star and $\hat{Y}$ in the direction of a true anomaly of $90^\circ$. This is a standard conversion of polar to Cartesian coordinates which gives

\begin{equation}
\bm{\hat{r}} = \begin{bmatrix}\cos f \\\sin f \end{bmatrix},
\end{equation}
and
\begin{equation}
\bm{\hat{f}}=\begin{bmatrix}-\sin f \\\cos f \end{bmatrix}.
\end{equation}
Equations~\ref{eq:vector_accel} and \ref{eq:vector_jerk} can then be expressed as
\begin{equation}
\label{eq:true_acceleration}
\frac{\ddot{\bm{r}}}{\lvert\ddot{\bm{r}}\rvert} = \begin{bmatrix}
-\cos f\\
-\sin f
\end{bmatrix},
\end{equation}
and
\begin{equation}
\label{eq:true_jerk_vector}
\frac{\dddot{\bm{r}}}{\lvert\ddot{\bm{r}}\rvert} = 
\begin{bmatrix}
\frac{2\dot{r}}{r}\cos f +\dot{f}\sin f \\\frac{2\dot{r}}{r}\sin f -\dot{f}\cos f 
\end{bmatrix}.
\end{equation}
Equation~\ref{eq:true_jerk_vector} can be expanded by finding expressions for $\dot{r}/r$ and $\dot{f}$; the former by differentiating $r$ in Equation~\ref{eq:r} with respect to time, and the latter by performing a similar time derivative on $f$, to obtain\footnote{This requires the fact that $\tan(f/2) = \kappa\tan(E/2)$ where $E$ is the eccentric anomaly and $\kappa \equiv \sqrt{1+e}/\sqrt{1-e}$. After taking the time derivative, $\dot{E}$ can be found by taking the time derivative of Kepler's equation and using the fact that the time derivative of the mean anomaly is $\dot{M} = 2\pi/P$.}

\begin{equation}
\label{rdotoverr}
\frac{\dot{r}}{r}=\frac{2\pi}{P}\frac{e\sin f\left(1+e\cos f \right)}{\left(1-e^2\right)^{3/2}},
\end{equation}
and
\begin{equation}
\label{fdot}
\dot{f} = \frac{2\pi}{P} \frac{\left(1+e\cos f \right)^2}{\left(1-e^2\right)^{3/2}}.
\end{equation}
Additionally, by combining Kepler's Third Law and Equations \ref{eq:r}, \ref{rdotoverr}, and \ref{fdot}, we can express the magnitude of acceleration and jerk, as well as the angle between them in full as

\begin{equation}
\label{eq:true_accel}
\lvert \ddot{\bm{r}}\rvert
 = \left(\frac{2\pi}{P}\right)^{4/3}\frac{G^{1/3}M_2}{\left(M_1+M_2\right)^{2/3}}\left(\frac{1+e\cos f}{1-e^2}\right)^2,
\end{equation}

\begin{equation}
\label{eq:true_jerk}
\frac{\lvert\dddot{\bm{r}}\rvert}{\lvert\ddot{\bm{r}}\rvert} = \frac{2\pi}{P}\frac{\left(1+e\cos f\right)}{\left(1-e^2\right)^{3/2}} \sqrt{4e^2\sin^2f+\left(1+e\cos f\right)^2},
\end{equation}
and
\begin{equation}
\label{eq:cos_true_angle}
\cos\theta=\frac{-2e\sin f}{\sqrt{4e^2\sin^2 f +\left(1+e\cos f\right)^2}}.
\end{equation}

\subsection{Projection onto the Sky Plane}

We can generate any orbit by rotating a general face-on orbit using the following extrinsic rotation matrix, $R$, using the $Z_\Omega X_i Z_\omega$ convention, where $\omega$ is the argument of periastron, $\Omega$ is the longitude of the ascending node, and $i$ is the inclination. We use the shorthand $S_u = \sin u$, $C_u = \cos u$ to condense the matrix as
\begin{equation}
    R = \begin{bmatrix}
C_\omega C_\Omega - S_\omega S_\Omega C_i & -S_\omega C_\Omega -C_\omega S_\Omega C_i & S_\Omega S_i\\
C_\omega S_\Omega + S_\omega C_\Omega C_i & -S_\omega S_\Omega + C_\omega C_\Omega C_i & -C_\Omega S_i\\
S_\omega S_i & C_\omega S_i & C_i
\end{bmatrix}.
\end{equation}

Then, $\bm{r}_o=R\bm{r}$ will give the vector $\bm{r}$ as observed from Earth that can be decomposed into perpendicular and radial components $\bm{r}_\perp$ and $\bm{r}_r$,
\begin{equation}
\bm{r}_o=\bm{r}_\perp+\bm{r}_r.
\end{equation}

In this transformation, the positive $X$-axis points north, the positive $Y$-axis points east, and the positive $Z$-axis points towards the observer. Note that the positive right ascension (RA) direction follows the positive $Y$-axis, and the positive declination (Dec) direction follows the positive $X$-axis in this convention. Therefore, we define $\hat{x}$ and $\hat{y}$ to represent the positive RA and Dec directions, respectively, and transform the vector according to $\hat{X}\rightarrow\hat{y}$ and $\hat{Y}\rightarrow{\hat{x}}$. This transformation simply swaps the x and y components to give

\begin{equation}
\label{eq:observed_acceleration}
\ddot{\bm{r}}_\perp = \begin{bmatrix}R_{21}\ddot{r}_{X} + R_{22}\ddot{r}_{Y}\\R_{11}\ddot{r}_{X} + R_{12}\ddot{r}_{Y}
\end{bmatrix},
\end{equation}
and
\begin{equation}
\label{eq:observed_jerk}
\dddot{\bm{r}}_\perp = \begin{bmatrix}R_{21}\dddot{r}_{X} + R_{22}\dddot{r}_{Y}\\R_{11}\dddot{r}_{X} + R_{12}\dddot{r}_{Y}
\end{bmatrix}.
\end{equation}

The $z$-components have been ignored but may be added in the case of any observed radial velocity variations. These are the accelerations and jerks projected onto the plane of the sky. The magnitudes of these projected accelerations and jerks can be written in a simple form using the fact that $\lvert\ddot{\bm{r}}_\perp\rvert^2=\lvert\ddot{\bm{r}}\rvert^2-\ddot{r}_{r}^2$ and $\lvert\dddot{\bm{r}}_\perp\rvert^2=\lvert\dddot{\bm{r}}\rvert^2-\dddot{r}_{r}^2$ to give
\begin{equation}
\label{eq:projected_accel}
\frac{\lvert\ddot{\bm{r}}_\perp\rvert^2}{\lvert\ddot{\bm{r}}\rvert^2} = 1-\sin^2 (i) \sin^2(f+\omega),
\end{equation}
and
\begin{equation}
\label{eq:projected_jerk}
\frac{\left|\dddot{\bm{r}}_\perp\right|^2}{\left|\dddot{\bm{r}}\right|^2}=1-\sin^2(i)\sin^2(f+\omega+\theta),
\end{equation}
where $\theta$ is the angle between the unprojected acceleration and jerk (those that occur within the orbital plane) calculated from Equation~\ref{eq:cos_true_angle}.
Additionally, the angle between the projected acceleration and jerk can be written as
\begin{equation}
\label{eq:cos_observed}
\cos(\theta_o)=\frac{\cos(\theta)-I_1I_2}{\sqrt{\left(1-I_1^2\right)\left(1-I_2^2\right)}},
\end{equation}
where
\begin{equation}
I_1 = \sin(i)\sin(f+\omega),
\end{equation}
and
\begin{equation}
I_2 = \sin(i)\sin(f+\omega+\theta).
\end{equation}
These equations provide projected accelerations and jerks at some instant in time. For consistency in this work and for comparisons, we choose this time to be the Gaia~DR3 epoch, which is J2016.0.

\subsection{Conversion to Physical Units}
The accelerations and jerks in the work above have units of meters per second squared and meters per second cubed; however, the astrometric accelerations and jerks reported in Astro-9 use milliarcseconds per year squared and milliarcseconds per year cubed. Since the parallax $\varpi$ in Astro-9 is in milliarcseconds, we only need to convert astronomical units to meters and years to seconds to obtain
\begin{equation}
\label{eq:wa_to_dpm}
\left|\ddot{\bm{r}}_{\perp}\right|=  1.50\times10^{-4}\,{\rm m}\,{\rm s}^{-2} \left( \frac{\left|\dot{\bm{\mu}}\right|}{1\,{\rm mas}\,{\rm yr}^{-2}} \right) \left( \frac{\varpi}{1\,{\rm mas}} \right)^{-1},
\end{equation}
and
\begin{equation}
\label{eq:wj_to_ddpm}
\left|\dddot{\bm{r}}_{\perp}\right|=  4.76\times10^{-12}\,{\rm m}\,{\rm s}^{-3} \left( \frac{\left|\ddot{\bm{\mu}}\right|}{1\,{\rm mas}\,{\rm yr}^{-3}} \right) \left( \frac{\varpi}{1\,{\rm mas}} \right)^{-1}.
\end{equation}
Note that Equations~\ref{eq:wa_to_dpm} and~\ref{eq:wj_to_ddpm} convert some measured astrometric accelerations and jerks in angular units to their physical units. Since these physical values are derived from astrometric quantities, we will still call them astrometric accelerations and jerks for the remainder of this work.

\subsection{Deriving System Parameters from Projections}
\label{sec:analyticalprojection}

In general, even stars with precise astrometric accelerations and jerks will have under-constrained orbits because there are four observables: $\dot{\mu}_{\alpha^*}$, $\dot{\mu}_{\delta}$, $\ddot{\mu}_{\alpha^*}$, and $\ddot{\mu}_\delta$, while there are seven unknowns: $M_2$, $P$, $e$, $f$, $i$, $\omega$, and $\Omega$. We explicitly assume that $M_1$ is well-measured by spectroscopy or SED fitting. We can alternatively minimally represent the observables as $\dot{\mu}=\left| {\dot{\bm \mu}} \right|$, $\ddot{\mu} = \left| \ddot{\bm \mu} \right|$, and the angle between them, $\theta_o$. Unfortunately, this still leaves the system underconstrained with three observables and six unknowns ($\Omega$ is removed, as can be seen in Equations~\ref{eq:projected_accel} and~\ref{eq:projected_jerk}). Two of the approximations allow estimates of orbital parameters using the observed $\dot{\mu}$, $\ddot{\mu}$, and $\theta_o$. We also provide an explicit formula for even the most general case.

\subsubsection{Inclined Eccentric Orbits}
Describing the equations for the inclined eccentric orbit case is excluded, since they are already provided in Equations~\ref{eq:projected_accel}-\ref{eq:cos_observed}, from which not even a one-parameter family of solutions can be obtained using only $M_1$, $\dot{\mu}$, $\ddot{\mu}$, and $\theta_o$. As an alternative, we present a combined form of Equations~\ref{eq:true_accel}-\ref{eq:cos_true_angle} that relate to the astrometric mass function:
\begin{equation}
    \frac{M_2^3}{\left( M_1 + M_2\right)^2} = \frac{P\lvert\ddot{\bm{r}}\rvert^6}{2\pi G\lvert\dddot{\bm{r}}\rvert^3\sin^3\theta}\left( 1-e^2 \right)^{3/2}.
    \label{eq:mass_function_eccentric_faceon}
\end{equation}
Note the similarity between the mass function in Equation~\ref{eq:mass_function_eccentric_faceon} and the spectroscopic mass function with radial velocity semi-amplitude $K$:
\begin{equation}
    \frac{{M_2^3}\sin^3i}{\left( M_1 + M_2\right)^2} = \frac{PK^3}{2\pi G}\left( 1-e^2 \right)^{3/2}.
\end{equation}
This is only to demonstrate the utility of astrometry, since in combination with spectroscopic orbits, these equations combine to give
\begin{equation}
\label{eq:simple_rv_and_astrometry}
\sin i=K\frac{\lvert\dddot{\bm{r}}\rvert\sin\theta}{\lvert\ddot{\bm{r}}\rvert^2}.
\end{equation}
Equation~\ref{eq:simple_rv_and_astrometry} reveals an independent test of the inclination of systems if given radial acceleration and jerk in addition to astrometric acceleration and jerk. It can also be used to estimate a lower bound on the inclination for Astro-9 systems with quadratic RV trends, since $\sin i$ is monotonic in $K$, and $K$ can be estimated for such systems.

\subsubsection{Face-On Eccentric Orbits}
 Once the inclination is removed, Equations~\ref{eq:projected_accel}-\ref{eq:cos_observed} are equivalent to Equations~\ref{eq:true_accel}-\ref{eq:cos_true_angle} since the orbital plane is observed.
 We may also remove $\Omega$ and $\omega$.\footnote{In this face-on eccentric case, rotational invariance allows us to ignore $\omega$, and $\Omega = 0$ by convention.} This leaves a system with four uncertain parameters: $M_2$, $P$, $e$, and $f$, and three observables: $\dot{\mu}$, $\ddot{\mu}$, and $\theta_o$, from which only a one-dimensional family of solutions can be derived. Note that in this case, $\lvert\ddot{\bm{r}}_\perp\rvert = \lvert\ddot{\bm{r}}\rvert$ and $\lvert\dddot{\bm{r}}_\perp \rvert = \lvert\dddot{\bm{r}}\rvert$, which are still related to $\dot{\mu}$ and $\ddot{\mu}$ through Equations~\ref{eq:wa_to_dpm}-\ref{eq:wj_to_ddpm}. If we square Equations~\ref{eq:true_accel}-\ref{eq:cos_true_angle}, derive $\tan^2\theta$, and define intermediate variables, 
 \begin{align}
    \mathcal{M} &\equiv \frac{M_2^3}{(M_1+M_2)^2}, & n &\equiv \frac{2\pi}{P}, \notag \\
    U &\equiv 1 + e \cos f, & V &\equiv 1 - e^2,
\end{align}
Equations~\ref{eq:true_accel}-\ref{eq:true_jerk} become

\begin{equation}
\label{eq:true_accel_sub}
\lvert \ddot{\bm{r}}\rvert^2
 = \frac{n^{8/3}{G^{2/3}\mathcal{M}^{2/3}}U^4}{V^4},
\end{equation}
and
\begin{equation}
\label{eq:true_jerk_sub}
\frac{\lvert\dddot{\bm{r}}\rvert^2}{\lvert\ddot{\bm{r}}\rvert^2} = \frac{n^2U^2}{V^{3}} \left(U^2+4\left[1-V-\left(U-1\right)^2\right]\right),
\end{equation}
while also obtaining
\begin{equation}
\label{eq:tan_true_angle_sub}
\tan^2\theta=\frac{U^2}{4\left[1-V-\left(U-1\right)^2\right]}.
\end{equation}

For brevity, we can then substitute Equation~\ref{eq:tan_true_angle_sub} into Equation~\ref{eq:true_jerk_sub} to obtain
\begin{equation}
\frac{\lvert\dddot{\bm{r}}\rvert^2}{\lvert\ddot{\bm{r}}\rvert^2} = \frac{n^2U^4}{V^{3}\sin^2\theta},
\end{equation}
which can be used with Equation~\ref{eq:true_accel_sub} to eliminate $U$ and $V$ in Equation~\ref{eq:tan_true_angle_sub}. Although it is possible to derive a cubic expression in $\mathcal{M}^{1/6}$, we opt to express the result in terms of $n$ since it can be cleanly isolated. It may also be convenient to define $\Psi\equiv G\mathcal{M}$ in a way reminiscent of the gravitational parameter. This finally gives
\begin{equation}
\label{eq:n_M_func}
n=\frac{\lvert\ddot{\bm{r}}\rvert^{3/4}}{8\Psi^{1/4}}\left[8-\frac{\lvert\dddot{\bm{r}}\rvert^2\Psi^{1/2}}{\lvert\ddot{\bm{r}}\rvert^{7/2}}\left(1+3\sin^2\theta\right)\right]^{3/2}.
\end{equation}
Note that Equation~\ref{eq:n_M_func} is also applicable in the inclined eccentric orbit case if it is additionally supplied with radial accelerations and jerks derived from radial velocity variations, which allows 3D accelerations, jerks, and angles to be calculated. In either case, Equation~\ref{eq:n_M_func} is only useful if an estimate of the orbital period can be made. On the other hand, since $P>0$, Equation~\ref{eq:n_M_func} requires
\begin{equation}
\label{eq:massupperbound}
G\mathcal{M} < \frac{64\lvert\ddot{\bm{r}}\rvert^{7}}{\lvert\dddot{\bm{r}}\rvert^{4}\left(1+3\sin^2\theta\right)^2}.
\end{equation}
Equation~\ref{eq:massupperbound} must hold throughout the orbit.
For systems that are in circular orbits or near either of the two apsides, $\theta\simeq90^{\circ}$, maximizing the denominator in Equation~\ref{eq:massupperbound}, potentially producing a valuable upper limit on $M_2$.

\begin{figure*}[t!]
\centering
\includegraphics[width=\columnwidth]{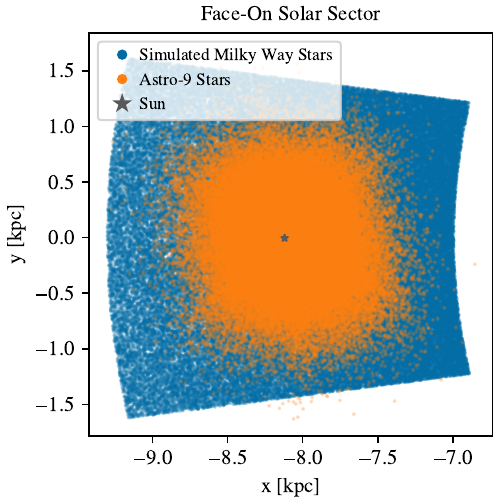}
\includegraphics[width=\columnwidth]{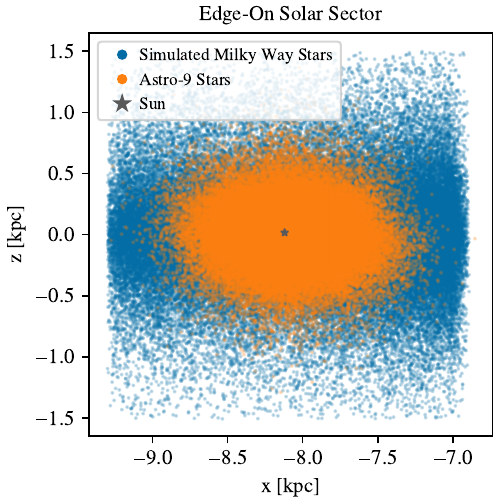}
\caption{View of Milky Way from above (left) and edge-on (right). We place randomly generated binaries (blue markers) at positions within a sector of the Milky Way centered on the Sun (dark gray star). This sector encloses the population of objects within the Gaia Astro-9 catalog of DR3 (orange markers).}
\label{fig:milky_way_positions}
\end{figure*}

\subsubsection{Face-On Circular Orbits}
\label{sec:face_on_circular}
In the simplest case, we remove both eccentricity and inclination. Equations~\ref{eq:true_accel}-\ref{eq:cos_true_angle} drastically simplify to
\begin{equation}
\label{eq:true_accel_circular}
\lvert \ddot{\bm{r}}\rvert
 = \left(\frac{2\pi}{P}\right)^{4/3}\frac{G^{1/3}M_2}{\left(M_1+M_2\right)^{2/3}},
\end{equation}

\begin{equation}
\label{eq:true_jerk_circular}
\frac{\lvert\dddot{\bm{r}}\rvert}{\lvert\ddot{\bm{r}}\rvert} = \frac{2\pi}{P},
\end{equation}
and
\begin{equation}
\cos\theta=0.
\end{equation}
We can directly plug Equation~\ref{eq:true_jerk_circular} into Equation~\ref{eq:true_accel_circular} to obtain

\begin{equation}
\label{eq:mass_circular}
\frac{GM_2^3}{(M_1+M_2)^2}=\frac{\vert\ddot{\bm{r}}\rvert^7}{\vert\dddot{\bm{r}}\rvert^4}.
\end{equation}
This is a cubic equation that can be solved for $M_2$. Additionally, Equation~\ref{eq:mass_circular} demonstrates a power-law relation between $\vert\ddot{\bm{r}}\rvert$ and $\vert\dddot{\bm{r}}\rvert$ when assuming sets of constant dark companion masses and constant luminous star mass. In log-space, this appears as a linear relationship:

\begin{equation}
\label{eq:log_circular_mass_equation}
\log(\lvert\ddot{\bm{r}}\rvert) = \frac{4}{7}\log(\lvert\dddot{\bm{r}}\rvert)+\frac{1}{7}\log\left[\frac{GM_2^3}{\left(M_1+M_2\right)^2}\right].
\end{equation}
As we demonstrate in subsequent sections, we can use Equation~\ref{eq:mass_circular} to quickly identify candidate systems that potentially contain a CO companion by estimating $M_1$ with the techniques mentioned in Section~\ref{sec:analyticalprojection} and calculating $\vert\ddot{\bm{r}}\rvert$ and $\vert\dddot{\bm{r}}\rvert$ from Equations~\ref{eq:wa_to_dpm}-\ref{eq:wj_to_ddpm}.

\section{Model Implementation}
\label{sec:model}
We start by adopting a model of the Milky Way stellar density from which to sample positions for our binaries. Then, we assign binary system parameters by drawing from three different orbital approximations: inclined eccentric, face-on eccentric, and face-on circular orbits. For a fourth binary distribution, we introduce a realistic binary population generated using the {\tt POSYDON} population synthesis code. Finally, for each of our populations, we apply an approximation of Gaia's selection function by using \texttt{gaiamock} to determine which systems would have been fitted with a 9-parameter variable acceleration solution.\footnote{The \texttt{gaiamock} fits are also in angular units, so their linear equivalents are calculated using Equations~\ref{eq:wa_to_dpm} and \ref{eq:wj_to_ddpm}.}

\subsection{The Milky Way Distribution}
\label{sec:milky_way}
To generate positions for binaries, we use a simple cylindrically symmetric model of the Milky Way stellar density that is exponentially decaying in the radial and axial directions with scale factors of $2.6\,\mathrm{kpc}$ and $0.3\,\mathrm{kpc}$, respectively. The Astro-9 catalog only contains stars within a distance of about $1.7\,\mathrm{kpc}$ from the Sun, so we only draw a wedge from the Milky Way density around the Solar neighborhood using inverse transform sampling of the exponential distribution. We use the Galactocentric coordinate system with cylindrical coordinates to describe this wedge, which corresponds to positions within $7\text{--}9.3\,\mathrm{kpc}$ from the Galactic Center, within $1.5\,\mathrm{kpc}$ of the Galactic plane, and within $170^\circ\text{--}190^\circ$ in azimuth. Following the default Galactocentric frame from \texttt{astropy v4.0}, we set the Sun to be $8.122\,\mathrm{kpc}$ from the Galactic Center, $20.8\,\mathrm{pc}$ above the Galactic plane, and at an azimuth of 180$^\circ$ in this coordinate system. Figure~\ref{fig:milky_way_positions} shows the resulting distribution of positions as blue points. The left panel shows the face-on projection of the distribution, while the right panel shows the edge-on projection. The orange points representing the Gaia Astro-9 objects are mostly contained within this wedge, which shows that our distribution of positions is sufficient to represent the observational distance constraints of the Astro-9 objects.

\subsection{Mock Populations}
\label{sec:mock_populations}
We generate three binary populations, one for each model of binary orbits: inclined eccentric, face-on eccentric, and face-on circular. The luminous star mass, $M_1$, is chosen to be $1\,M_\odot$ to represent a Sun-like star, while $M_2$ is a dark companion; we separately test dark companion masses of $0.01$, $0.1$, $1$, and $10\,M_\odot$.  The orbital period, $P$, is chosen to follow a log-uniform distribution ranging from $10^2-10^4\,\mathrm{days}$. We treat each object as a point mass and assume that any flux originates strictly from $M_1$. Furthermore, no binary evolution is incorporated. Then, each population is generated as follows:

I) For our most general population, inclined eccentric binaries, the orbital eccentricity $e$ follows a thermal distribution ($p(e)=2e$). The mean anomaly $M$, argument of periastron $\omega$, and longitude of ascending node $\Omega$, are drawn uniformly from $[0, 2\pi)$ as no preference is expected. We obtain the inclination $i$ by drawing $\cos i$ uniformly from $[-1, 1]$ due to the isotropic distribution of the orbits from the perspective of Earth.

II) The next population, face-on eccentric binaries, only changes from (I) by setting the inclination to zero, which makes the longitude of the ascending node degenerate, so we set it to zero by convention.

III) Our simplest population, face-on circular binaries, simplifies (II) by setting $e=0$, for which the argument of periastron becomes degenerate, so we set it to zero by convention.

\begin{figure*}[!ht]
\centering
\includegraphics[width=\columnwidth]{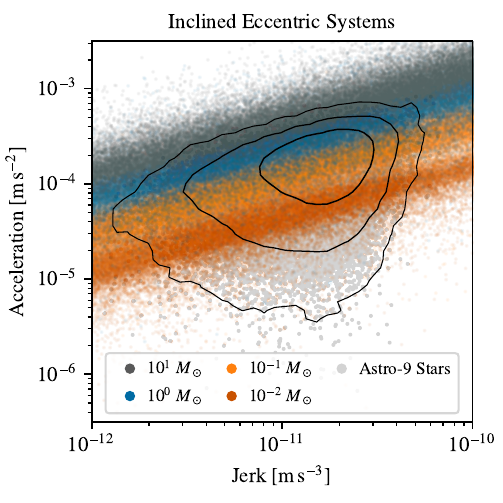}
\includegraphics[width=\columnwidth]{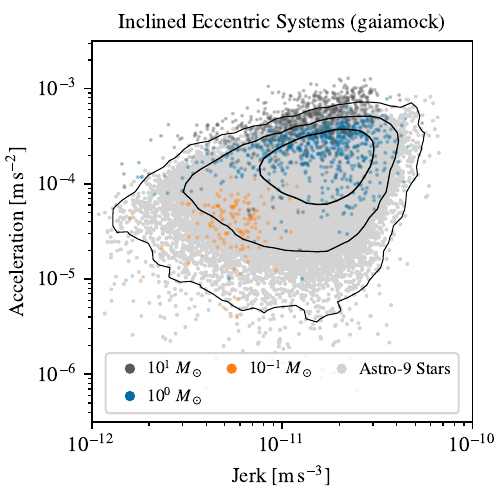}
\caption{The randomly generated population of inclined eccentric binaries, along with the Gaia Astro-9 sample. Left panel: projected accelerations and jerks obtained from Equation~\ref{eq:projected_accel} and Equation~\ref{eq:projected_jerk}. The samples follow a power law, but with a wide spread. Clear banding shows that systems with specific dark companion masses mainly occupy specific regions of the plot. Right panel: Astrometric accelerations and jerks after processing the sample through {\tt gaiamock}. Even though only a subset of our mock samples are given a 9-parameter solution, systems with specific dark companion masses still tend to occupy different regions in this plot. Included are black contours to discern the distribution of Astro-9 stars from our mock populations. From innermost to outermost, these contours enclose 68\%, 95\%, and 99.7\% of the Astro-9 stars.}
\label{fig:eccentric_inclined_both}
\end{figure*}

\subsection{\texttt{POSYDON} Mock Population}
\label{sec:POSYDON}
In addition to the three previous analytic populations, we generate a fourth population using version 2.3.1 of the binary population synthesis code \texttt{POSYDON} \citep{2023ApJS..264...45F, 2025ApJS..281....3A}. \texttt{POSYDON} generates large populations of stellar binaries by interpolating precomputed grids of \texttt{MESA} \citep{2011ApJS..192....3P, 2013ApJS..208....4P, 2015ApJS..220...15P, 2018ApJS..234...34P, 2019ApJS..243...10P, 2023ApJS..265...15J} binary simulations. These grids are stitched together to form the complete evolution of binaries from ZAMS to double compact object formation. Since it includes prescriptions to handle binary evolution processes, including mass transfer and supernovae (SNe), \texttt{POSYDON} provides a population of BH binaries following state-of-the-art stellar modeling.

We use \texttt{POSYDON} to evolve a population of $500,000$ stellar binaries at solar metallicity. The masses of the primaries follow the Kroupa initial mass function \citep{2001MNRAS.322..231K} between $15\,M_\odot$ and $150\,M_\odot$, while a uniform mass ratio distribution sets the masses of the secondaries between $0.1\text{--}2\,M_\odot$. We initialize the binary orbits as circular with orbital periods drawn from an extended \citet{2012Sci...337..444S} period distribution described in \citet[][Appendix A]{2021A&A...647A.153B}. To determine the explodability and outcome of SNe, we use the delayed SN-engine model from \citet{2012ApJ...749...91F}, with natal BH kicks drawn from a Maxwellian distribution with a velocity dispersion of $265\,\mathrm{km\,s^{-1}}$.

After evolving the population, we select only detached binaries composed of a black hole and a non-degenerate companion. These systems are given random orientations according to the distributions in Section~\ref{sec:mock_populations}. Finally, we compute absolute $G$ magnitudes using \texttt{MSG}  \citep{Townsend2023}, which utilizes multivariate interpolation within pre-calculated spectral grids to produce spectra and photometric colors \citep[for details, see][]{2026arXiv260613351K}. In this work, we use the C3K spectral library of \citet{2016ApJ...823..102C} that is based on the ATLAS12 models of \citet{1970SAOSR.309.....K, 1993sssp.book.....K}.

\begin{figure*}[!ht]
\centering
\includegraphics[width=\columnwidth]{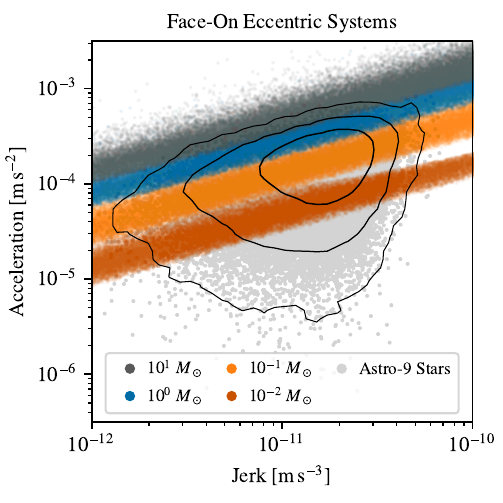}
\includegraphics[width=\columnwidth]{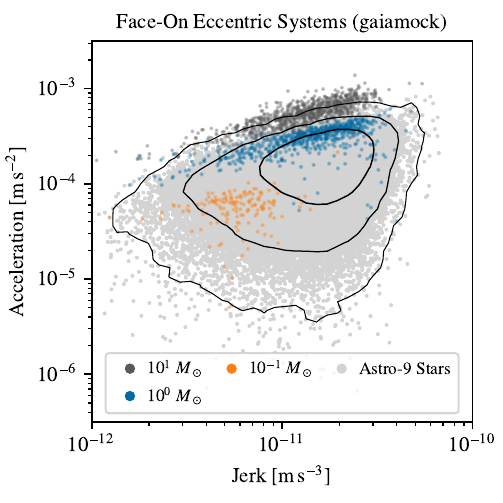}
\caption{Left: Accelerations and jerks obtained from Equation~\ref{eq:true_accel} and Equation~\ref{eq:true_jerk} (since the orbital plane is observed). As expected, systems with different mass companions are clearly separated, with more massive dark companions causing higher accelerations for any particular value of the jerk. The spread around the power law described by Equation~\ref{eq:log_circular_mass_equation} is tighter compared with the distribution in Figure~\ref{fig:eccentric_inclined_both}, but is still significant. Right: Astrometric accelerations and jerks from \texttt{gaiamock}. As before, we see that the number of systems fit with a 9-parameter solution is significantly less than total number of systems generated. The prominences below the dense linear cores are more apparent now which occur at higher jerks. Included are black contours to discern the distribution of Astro-9 stars. These contours enclose 68\%, 95\%, and 99.7\% of the Astro-9 stars.}
\label{fig:eccentric_face_on_both}
\end{figure*}

\subsection{Detectability}
\label{sec:detectability}
As the final step in our process, we use \texttt{gaiamock} \citep{2024OJAp....7E.100E} to incorporate the detectability of each generated binary system. This code forward-models Gaia's selection function to determine astrometric solutions of synthetic observations. It attempts to fit 5-parameter, 9-parameter, 7-parameter, and then full orbital solutions to observations, just as Gaia's astrometric pipeline. As input, we use each of our generated systems' Galactic position (in right ascension, declination, and distance) and orbital parameters (from the distributions described in Sections~\ref{sec:milky_way} and \ref{sec:mock_populations}). The proper motions are assumed to be zero. For the \texttt{POSYDON} population, we use the absolute $G$ magnitude it calculates, while for our other three populations we assume that the absolute $G$ magnitude of every luminous star is that of the Sun, $M_{G,\odot}=4.68$, as calculated by \citet{10.1093/mnras/stz1397}. The apparent $G$ magnitudes are calculated using the distance modulus with a simplified linear dust extinction of $1\,\mathrm{mag\,kpc^{-1}}$ \citep[although we note that the actual extinction is a complex function of both sky position and distance;][]{2019ApJ...887...93G}. Any system not immediately fit with a 9-parameter astrometric solution is ignored. This provides us with a sample of only 9-parameter astrometric solutions for each mock population.

\section{Results}
\label{sec:results}
Here we provide both the projected accelerations and jerks for each population and their respective astrometric fits from \texttt{gaiamock}. As mentioned in Section~\ref{sec:analyticalprojection}, the equations provide values at one moment in time, so the projected acceleration and jerk for each system are the instantaneous values that occur at some true anomaly, while the \texttt{gaiamock} fits are derived from positions observed within a 34-month baseline.

\subsection{Inclined Eccentric Orbits}
In the left panel of Figure~\ref{fig:eccentric_inclined_both}, we show the distributions of projected accelerations and jerks from our four populations, each having different dark companion masses: $10\,M_{\odot}$ (black points; characteristic of a black hole), $1\,M_{\odot}$ (blue points; similar to a neutron star or white dwarf), $0.1\,M_{\odot}$ (orange points; such as a small M dwarf), $0.01\,M_{\odot}$ (brown points; such as a brown dwarf or giant planet). The wide range of orbital periods, eccentricities, orbital phases, and inclinations leads to a broad scatter of points that extend beyond the plot limits, far outside the distribution of points obtained from the Gaia Astro-9 sample. Nevertheless, inclined eccentric systems produce projected accelerations and jerks that roughly follow a power-law distribution, as expected from Equation~\ref{eq:log_circular_mass_equation}. The separation between populations indicates that even after accounting for inclination, a clear trend with mass exists, suggesting that the majority of dark companion masses can be estimated on a statistical basis using the projected acceleration and jerk of the luminous star.

\begin{figure*}[!ht]
\centering
\includegraphics[width=\columnwidth]{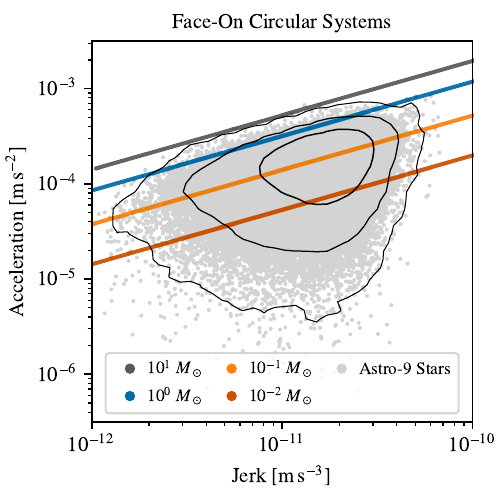}
\includegraphics[width=\columnwidth]{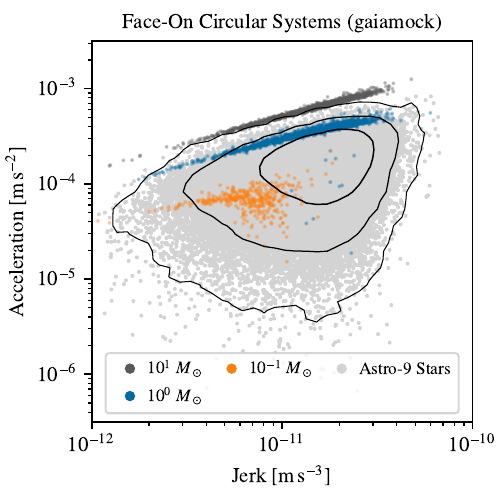}
\caption{Left: Accelerations and jerks obtained from Equation~\ref{eq:true_accel_circular} and Equation~\ref{eq:true_jerk_circular}. The effect of eccentricity and inclination are gone leaving behind clear linear distributions for each dark companion mass. While lower mass objects can cause larger accelerations, they are distinguished from higher mass companions by differences in the jerks they impose. Right: Astrometric accelerations and jerks from \texttt{gaiamock}. The same power-law trend is apparent. However, more massive and longer-period systems tend to have astrometric fits that more closely follow the projected values, while the shorter-period systems tend to deviate from the projected values, which suggests that astrometric fits to systems that have orbital periods shorter than the DR3 time baseline are likely to be inaccurate and thus less useful for predictions. This is most notably seen in the high-jerk regime for systems with a $0.1\,M_\odot$ companion (orange points). Included are black contours to discern the distribution of Astro-9 stars. From innermost to outermost, these contours enclose 68\%, 95\%, and 99.7\% of the Astro-9 stars.}
\label{fig:circular_face_on_both}
\end{figure*}

In the right panel, we see that the range of projected accelerations and jerks is significantly reduced (shown in linear units), which highlights the strong selection bias present. Some systems also tend to have their astrometric accelerations systematically underestimated compared to their projected values. The distributions still span a range of projected jerks, but now there are clear prominences extending out from the dense linear cores. Despite this, we can still see that systems with different dark companion masses mostly separate, with $0.1\,M_\odot$ dark companions occupying the low projected acceleration and low projected jerk regime, and systems with $10\,M_\odot$ dark companions occupying the high projected acceleration regime. An important note is that this result suggests that astrometric accelerations and jerks are rarely larger than their projected counterparts. For example, an astrometric acceleration and jerk characteristic of a $10\,M_\odot$ dark companion is unlikely to be confused with a system having a $0.1\,M_\odot$ dark companion. We also note that essentially none of our systems with companion masses of $0.01\,M_\odot$ are detectable in this dataset.

\subsection{Face-On Eccentric Orbits}
Compared with the inclined eccentric binaries in Figure~\ref{fig:eccentric_inclined_both}, we see that the distributions in the left panel of Figure~\ref{fig:eccentric_face_on_both} still exhibit the characteristic power-law stripes that span the plot axes. However, in the face-on case, the distributions are narrower, with less overlap between systems with different dark companion masses at different orders of magnitude. 
After accounting for detectability using \texttt{gaiamock}, the right panel of Figure~\ref{fig:eccentric_face_on_both} shows a trend similar to the right panel of Figure~\ref{fig:eccentric_inclined_both}. However, when orbits are assumed to be face-on, the distributions are tighter with less overlap, suggesting that the companion mass can be more reliably estimated with acceleration and jerk alone. Finally, we note that the astrometric accelerations for systems with higher astrometric jerks are still significantly underestimated compared to the projected accelerations. Deviations from the simple power law described by Equation~\ref{eq:mass_circular} are most visually apparent in the samples with $0.1\,M_\odot$ and $1\,M_{\odot}$ dark companion masses, which show prominences below the dense linear cores.

\subsection{Face-On Circular Orbits}
In the left panel of Figure~\ref{fig:circular_face_on_both}, we show the accelerations and jerks that were generated under the assumption of face-on circular orbits. Following Equation~\ref{eq:log_circular_mass_equation}, the stratification between systems with different dark companion masses is now obvious. Although systems can be degenerate using acceleration alone, they become distinct when jerk is included. This means that any face-on circular system can, in theory, have each parameter determined using only acceleration and jerk.

In the right panel of Figure~\ref{fig:circular_face_on_both}, we show these same orbits as processed by \texttt{gaiamock} to determine their observability. As was the case with the face-on eccentric binaries, the astrometric values tend to be distributed along the power laws shown in the left panel. The resulting distributions also closely resemble the dense cores of the distributions shown in Figures~\ref{fig:eccentric_inclined_both} and~\ref{fig:eccentric_face_on_both}. There is also a clear deviation from the projected values at shorter orbital periods, most notably for the systems with $10^{-1}\,M_\odot$ dark companions. This suggests that the astrometric fits are absorbing unmodeled higher-order terms.

\subsection{POSYDON Black Hole Sample}
\begin{figure}
\centering
\includegraphics[width=\columnwidth]{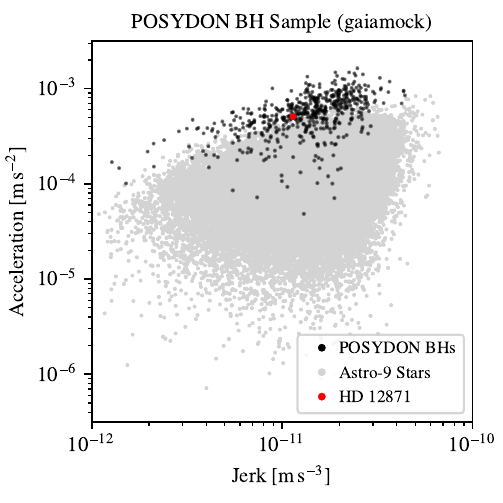}
\caption{POSYDON sample of BH binaries processed with \texttt{gaiamock}. The distribution is similar to that of the inclined and eccentric orbits shown in Figure 
\ref{fig:eccentric_inclined_both} where the greatest concentrations of systems tend to lie along the linear distributions defined by the face-on circular orbits. While a few systems lie among the less massive systems, having lower astrometric accelerations, most are found in the upper region. The slightly enlarged red marker shows the location of HD~12871 on this plot.}
\label{fig:posydonbhsample}
\end{figure}

Finally, we provide the result of BH binaries created using \texttt{POSYDON} and processing with \texttt{gaiamock}. Unlike the previous three mock populations, this \texttt{POSYDON}-generated sample accounts for binary evolution, enabling inspection of a more realistic population. Figure~\ref{fig:posydonbhsample} shows the astrometric accelerations and jerks of this sample, comprising luminous stars between $0.6\text{--}1.0\,M_\odot$ and BH companions between $10\text{--}20\,M_\odot$. The appearance is similar to the $10\,M_\odot$ distribution in the right panel of Figure~\ref{fig:eccentric_inclined_both}, with the majority of systems lying in the unmistakable upper region where we suspect compact object companions to occupy. Some systems have lower astrometric accelerations, lying in the region that is indicative of lower mass companions, though these are rare. Given their overlap with many other Gaia binaries and limited resources for follow-up campaigns, we expect that such high-mass systems with lower astrometric accelerations will be extremely challenging to distinguish from low-mass systems with similar values and thus cannot be prioritized for follow-up based on astrometric acceleration and jerk alone.

Compared with the background distribution of Astro-9 stars, many of our BH-hosting systems shown in Figure~\ref{fig:posydonbhsample} exhibit larger astrometric accelerations for a given astrometric jerk. Certainly, the \texttt{POSYDON} model is an imperfect representation of the Milky Way sample, and model improvements may lead to differences in the BH distribution in this space. Nevertheless, given that the vertical position of the distribution is primarily affected by dark companion masses (as seen, for example, from the power law distribution in Figure~\ref{fig:circular_face_on_both}), differences between the \texttt{POSYDON} BH and Astro-9 distributions imply that BH-hosting binaries are likely to be rare and with relatively low-mass BHs within Astro-9.

We note that the distribution of black markers shown in Figure~\ref{fig:posydonbhsample} is therefore meant to provide a prediction for the region where BH binaries are expected to reside rather than a prediction for the number expected to be detectable. Population synthesis predictions for the overall rate of such binaries are highly uncertain, not just because of binary evolution uncertainties, but also because the formation rate of initial binaries with extreme mass ratios is observationally ill-constrained. For example, Gaia BH1 and BH2, whose puzzling origins have many hypotheses in the literature, both consist of a $\simeq10\,M_{\odot}$ BH with a $\simeq1\,M_{\odot}$ companion, which implies an initial mass ratio $q\lesssim0.03$.

\section{Finding Compact Objects}
\label{sec:discussion}
When searching for compact objects in binaries, it is important to be aware of false-positives, as well as the best place to look. Here, we describe potential sources of contamination (systems that can appear as hosting a BH companion), as well as the best place to look according to the \texttt{gaiamock} fits of the \texttt{POSYDON} population. We also include a discussion of HD~12871, a historical object-of-interest, and \texttt{gaiamock} fits of different realizations of Gaia BH3.

\subsection{Sources of Contamination}
\label{sec:contamination}
The Gaia DR3 catalog of non-single stars contains 91,268 Astro-9 objects. Since the general problem of an inclined, eccentric orbit is under-constrained, we cannot solve exactly for the masses and periods of each system in this sample. In principle, one could use the measured values of $\dot{\mu}$, $\ddot{\mu}$, and $\theta_o$, along with the assumption of a face-on eccentric orbit, to obtain a one-dimensional relation between the orbital period and dark companion mass. \citet{2023ApJ...946..111A} showed that in the case of accelerating sources, one can place a lower limit on the dark companion mass by setting a minimum orbital period to be the time baseline of the catalog, 34 months for DR3. Similarly, one could apply the same constraint to Astro-9 sources, since in general, systems showing the same astrometric acceleration and jerk have a trend of higher companion masses for longer orbital periods. 

However, as \citet{2024OJAp....7E.100E} point out, the presence of a short-period tail complicates the ability to estimate companion masses. Our work focuses on the distribution of compact objects in binaries within the Astro-9 sample, so the orbital periods are less relevant for determining candidate systems. Although systems with orbital periods less than the 34-month baseline of Gaia DR3 can sometimes be fit with a 9-parameter solution according to \texttt{gaiamock}, we have found that the estimated orbital parameters of these systems are unreliable. This is because the astrometric accelerations and jerks of these systems typically do not match the projected values assumed in the derivations in Section~\ref{sec:methods}. In general, for a given astrometric acceleration and jerk, it is challenging to determine which systems have long-period orbits with accurate fits and which systems have much shorter periods\footnote{Similar to, or less than, the 34-month time baseline of DR3} with inaccurate fits. Future Gaia data releases that provide time-series astrometric observations may afford a careful distinction between the two possibilities. For now, we are limited to accepting that such short-period systems provide a significant source of contamination in the Astro-7 and Astro-9 catalogs. \citet{Simon_2026} contains examples of such systems among their data, where some of their CO candidate systems have Astro-7 and Astro-9 solutions.

While our analysis assumes that the companion star is dark and contributes no significant emission, binaries with non-degenerate companions form additional sources of contamination. In such cases, Gaia observes the photocenter of the system. Since the photocenter lies somewhere between the two stars and the timescale of variation remains the same, the photocenter motion will lead to underestimated astrometric accelerations and jerks. The Milky Way is replete with non-degenerate binaries, some of which will be eccentric and will have had orbital phases near periastron at the J2016.0 reference epoch where both $\left| \ddot{\bm{r}} \right|$ and $\left| \dddot{\bm{r}} \right|$ in Equations~\ref{eq:true_accel} and \ref{eq:true_jerk} may be instantaneously large, which can be mistaken as being due to a massive object. That is to say, an Eddington bias may lead to sources with lower-mass companions appearing to host massive, dark companions. Furthermore, our figures do not include error bars for astrometric accelerations and jerks.

Along this same vein, but with multiple-star systems, the tight inner binary of a triple system or even the inner hierarchy of higher order systems can also contaminate Astro-9, an example of which we will discuss in Section~\ref{sec:HD12871}. If the combined flux of the inner binary is much less than the flux of the luminous outer star, the inner binary effectively acts as a single massive and dark object. In this work we do not model triple or higher order systems, but instead note that for sufficiently faint and compact inner hierarchies, we expect the astrometry of a luminous outer star to appear as only having a single massive companion. In contrast, \citet{10.1093/mnras/stad3294} suggests that Gaia BHs in open clusters often have tertiary stars. We leave the details of detecting triple systems within astrometric binary data for future work. However, we point out that such triple systems, whether compact object hosting or not, must be stable for long times to be detectable in substantial numbers in the Milky Way field. Accounting for such stability may lead to additional constraints on the nature of hierarchical systems.

Regardless of any contamination, even in regions that overlap significantly with the predictions from our models (where massive companion signatures often appear), we expect BH-hosting binaries to be an exceptional rarity.\footnote{As one can see around the $10\,M_\odot$ points on the \texttt{gaiamock} side of any figure, the actual number of objects within the overlap with Astro-9 is quite small.} Our goal throughout this work is to identify regions that are most likely to host BHs to optimize follow-up campaigns rather than to assert the nature of any individual system. Given the checkered history of identifying BH binaries, one should tread lightly when considering even ideal cases. Nevertheless, our approach using the Astro-9 catalog allows for a robust way to optimize follow-up campaigns for long-period systems.

\subsection{Compact Object Candidates by Relative Probability}
\label{sec:candidates}
High astrometric accelerations are directly related to higher masses and shorter periods, but our results show that it is not strictly systems with the highest astrometric accelerations that are most likely to host compact object companions. We find that the astrometric jerk provides a distinguishing factor for objects of similar astrometric accelerations. Using the face-on circular orbits as a reference, stars with an astrometric acceleration of $10^{-4}\,\mathrm{m\,s^{-2}}$ can easily have companions around $0.01\text{--}10\,M_\odot$ but are separated in astrometric jerk by orders of magnitude. Following Equation~\ref{eq:mass_circular}, the most straightforward approach to optimize observational follow-up is to identify systems with large values of the combined quantity $\lvert\ddot{\bm{r}}\rvert^7\,\lvert\dddot{\bm{r}}\rvert^{-4}$.

\begin{figure}[h]
\centering
\includegraphics[width=\columnwidth]{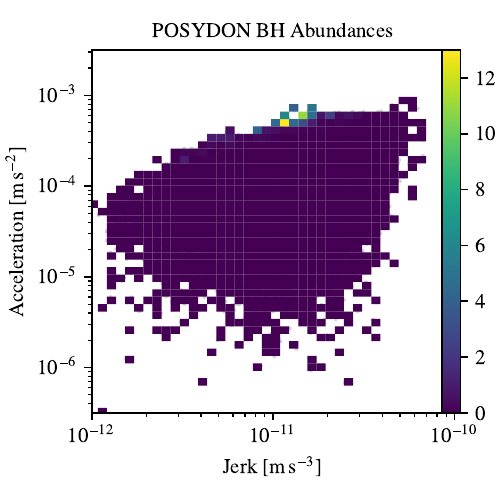}
\caption{Density map of the number of \texttt{POSYDON} BH systems with compact object masses between $10\text{--}20\,M_{\odot}$ relative to Gaia~DR3 Astro-9 objects. In the upper region, there is a clear abundance of BH systems. The vast majority of the parameter space indicates a low probability of Astro-9 objects hosting compact object companions, allowing observational follow-up campaigns to focus their efforts on a relatively small list of targets. We note that this map does not directly prove whether any particular Astro-9 object is a BH system, but rather if it is a reasonable object to follow up.}
\label{fig:BHabundances_posydon}
\end{figure}

While maximizing this quantity provides a broad expectation for which systems are most likely to host BH companions, our model predictions from \texttt{POSYDON} furnish an alternative, quantitative approach. Adopting the assumption that BH binaries are a very rare subset of the overall sample of Astro-9 objects, we can approximate the likelihood of a BH companion as the ratio between the number of BH binaries with a particular astrometric acceleration and jerk and the number of overall sources in the Astro-9 catalog with those same values, realized as
\begin{equation}
\label{eq:probabilityBH}
P_{\mathrm{BH}} \approx \frac{N_{\mathrm{BH}}(\lvert\ddot{\bm{r}}\rvert, \lvert\dddot{\bm{r}}\rvert)}{ N_{\mathrm{Gaia}}(\lvert\ddot{\bm{r}}\rvert, \lvert\dddot{\bm{r}}\rvert)}.
\end{equation}
We use the \texttt{POSYDON} BH sample described in Section~\ref{sec:POSYDON} for binary predictions because it naturally incorporates binary evolution along with a proper implementation of the initial binary distributions. This sample is not normalized to the overall rate of BH binaries expected in the Solar Neighborhood. However, for the purposes of identifying the sources most likely to host BH companions, the difference is a normalizing constant that can be ignored.

Figure~\ref{fig:BHabundances_posydon} shows a density map of the ratio between the number of BH systems generated by \texttt{POSYDON} and the number of Astro-9 systems using $[100 \times 100]$ binning over the Astro-9 systems.\footnote{In particular, over the minimum and maximum astrometric acceleration and jerk (in physical units) of the Astro-9 systems: $[\dddot{\mathbf{r}}_{\mathrm{min}},\dddot{\mathbf{r}}_{\mathrm{max}}]\times[\ddot{\mathbf{r}}_{\mathrm{min}},\ddot{\mathbf{r}}_{\mathrm{max}}]$. This ignores \texttt{POSYDON} BH systems that are not within the Astro-9 bounds, allowing for a direct comparison.} In the vast majority of parameter space, the overlap is sparse. This is a good thing, as it allows focused efforts on the clear concentration of systems with astrometric accelerations around $10^{-4}\text{--}10^{-3}\,\mathrm{m\,s^{-2}}$ and astrometric jerks near $10^{-11}\,\mathrm{m\,s^{-3}}$. The density plot also demonstrates a distinct hotspot at about $10^{-3.5}\,\mathrm{m\,s^{-2}}$ and $10^{-10.95}\,\mathrm{m\,s^{-3}}$.
These results highlight the same findings from our modeled systems: BH candidates are likely to be found with high astrometric accelerations but distinguished from non-BH systems statistically by their astrometric jerks.

\subsection{HD~12871}
\label{sec:HD12871}
Although observational follow-up for the most promising compact object candidates in the Astro-9 catalog is outside the scope of this work, there are systems that have partial orbital solutions in the literature, determined through radial velocity campaigns. Here we highlight one of these, HD~12871. HD~12871 is found within the region that we determine to be the most likely to contain BH candidates, shown as the red marker in Figure~\ref{fig:posydonbhsample}. This system has an RV curve calculated by \citet{1994Obs...114...45G}, who found that the system has an orbital period of $1370.6\,\mathrm{d}$, an eccentricity of 0.197, and a mass function of $0.369\,M_\odot$. Assuming that the luminous star has a mass of $1\,M_\odot$, the hidden component is expected to be at least $1.22\,M_\odot$. \citet{1994Obs...114...45G} speculated on the possibility that HD~12871 hosts a neutron star or black hole companion; however, these authors dismissed the idea, instead preferring to explain the non-detection of the secondary star as a pair of late K-dwarfs whose spectra could hide within that of the primary. Long-term follow-up observations by \citet{10.1093/mnras/stu1838} found an additional component to the spectrum that varies on an approximately month-long timescale, making HD~12871 at least a triple system or even higher order (J.L.~Halbwachs, private communication).

Despite the fact that HD~12871 is unlikely to host a compact object companion, the archival radial velocity solution confirms the validity of our method: optimizing for astrometric acceleration and jerk can aid in identifying binary systems with large mass functions. Since HD~12871 is likely a triple system or a higher-order system, there are possibly more systems in the same region that are also triples. On the other hand, the multiple-order explanation works only for sources with a relatively lower predicted candidate dark mass, as mentioned in Section~\ref{sec:contamination}. Perhaps in the case of HD~12871 the $\simeq1\,M_{\odot}$ luminous star could outshine two $\simeq0.6\,M_{\odot}$ K dwarfs in a tight binary, but it would be challenging to explain a $10\,M_{\odot}$ companion as a compact system of many low-mass stars. Therefore, while there is a clear potential for triple contamination, we may be able to filter out systems where the faint companion mass is unreasonably large. Exactly where this mass limit lies requires a more detailed analysis and will vary depending on the specific parameters of any particular system.

\subsection{Gaia~BH3}
\label{sec:known-systems}
\begin{figure}[t]
\centering
\includegraphics[width=\columnwidth]{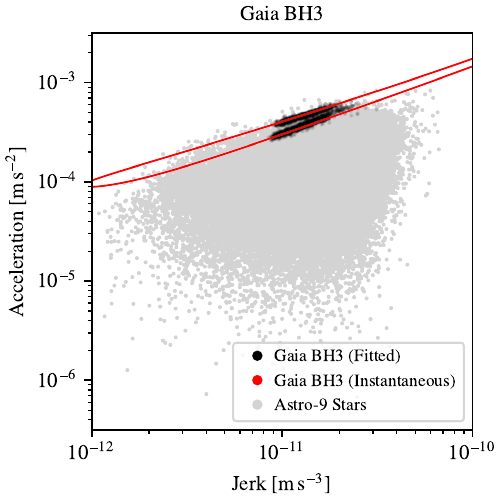}
\caption{Comparison of the projected acceleration and jerk of Gaia~BH3 (red line) to the astrometric fits from \texttt{gaiamock}, assuming various periastron times. For certain times, \texttt{gaiamock} predicts Gaia~BH3 would have been included in the Astro-9 catalog, so we expect its exclusion from was due to bad luck. Since its 11.6-year orbital period is significantly longer than the time baseline of Gaia~DR3, the astrometric accelerations and jerks closely follow the projected values, making the approximations taken in deriving our results applicable. The fact that mock fits of the Gaia~BH3 orbit overlap with the region of interest among the Astro-9 stars suggests that there may be other BHs in the Astro-9 catalog that could be confirmed with future Gaia releases.}
\label{fig:obsGaiaBH3}
\end{figure}

Gaia~BH3 was discovered in Gaia~DR4 pre-release data by \citet{2024A&A...686L...2G} as hosting a $33\,M_{\odot}$ BH in an 11.6-year orbit with a $0.76\,M_{\odot}$ luminous star. Although this system was not listed in the Gaia~DR3 Astro-9 catalog, we find that its exclusion is due to bad luck, since from $10^4$ realizations of Gaia BH3, only a few hundreds were fit with an Astro-9 solution. In Figure~\ref{fig:obsGaiaBH3}, we test the observability of Gaia~BH3 using \texttt{gaiamock} for different orbital phases. We utilize the system's published orbital solution to calculate the projected acceleration and jerk (red line) of the luminous star along its entire orbit and compare to its astrometric acceleration and jerk (black markers) from mock observations of the system with various periastron times.\footnote{As input to \texttt{gaiamock}, varying the time of periastron varies the 34-month segment of the orbit observed.} For a range of periastron times, the system is fitted with an astrometric acceleration and jerk that accurately follows the projected values. This accuracy is due to the system's long orbital period, which is much longer than Gaia~DR3's baseline, making the approximations taken in deriving our results applicable to Gaia~BH3. Indeed, the system's astrometric acceleration and jerk place it firmly in the region where compact object companions are likely to be. This will be confirmed by the data coming soon from Gaia~DR4.

\section{Conclusions}
\label{sec:Conclusion}
We used Newton's Law of Gravitation to find the orbital acceleration and jerk of an arbitrary binary system composed of a luminous star and a compact object and projected them onto the plane of the sky. We randomly generated a large sample of systems with a range of orbital periods, orbital orientations, and dark companion masses under three separate orbit approximations: circular and face-on, eccentric and face-on, and eccentric and inclined. We then randomly placed these systems in the Solar Neighborhood and calculated their detectability and observational biases within Gaia astrometric data using \texttt{gaiamock}. We additionally generated a random sample of BH-hosting binaries using \texttt{POSYDON}, which incorporated realistic binary evolution physics.

We found that systems with large companion masses occupy a specific region in the projected acceleration--jerk parameter space. Even though the assumption of a face-on, circular orbit is rather simplistic, it provides an estimate for the dark companion mass that is largely consistent with the more realistic case of an eccentric, inclined orbit, even when accounting for observational biases. While we expect BH binaries to be rare and contamination to be significant, our results point to a simple approach to optimize the search for systems most likely to host massive dark companions by identifying systems with a large value for the quantity $\lvert\ddot{\bm{r}}\rvert^7\,\lvert\dddot{\bm{r}}\rvert^{-4}$.
Alternatively, our \texttt{POSYDON} population synthesis predictions can be combined with the Gaia Astro-9 catalog to find the regions in the astrometric acceleration--jerk space where the ratio between the number of predicted BHs and the number of sources in the Astro-9 catalog is maximized.

While observational follow-up of individual systems is outside the scope of this work, we provided an example system from the Astro-9 catalog, HD~12871, as both a demonstration of our method and a cautionary tale. A complete radial velocity curve from the literature leads to a large mass function that implies a massive companion, as expected from its placement in the astrometric acceleration--jerk space. However, high-resolution spectroscopic follow-up found contributions from at least one additional component in the system, suggesting that it is likely a triple or higher-order system.

With its enhanced precision and time baseline nearly double that of DR3, the upcoming release of Gaia DR4 promises to be a significant enhancement over the currently available DR3 data. If any system in the DR3 Astro-9 sample is found within DR4 to host a BH companion, we can confirm our approach. By providing individual measurements of positions and, for sufficiently bright stars, their radial velocities, the Gaia DR4 data offer an additional advantage. If the radial velocity fits show any curvature, then even for systems with incomplete orbits in DR4, with all three dimensions of acceleration and jerk measured, we can de-project the orbits and compare against our models adopting the eccentric face-on orbital assumption. Finally, we point out that if our approach is confirmed using DR4, we can use it to optimize follow-up observations for the even longer-period systems exhibiting incomplete orbits in DR4; essentially a whole new set of systems with 9-parameter solutions. This is perhaps the best and only way to identify massive compact object binaries that have orbital periods of 10 or more years.

\begin{acknowledgments}

The authors acknowledge UFIT Research Computing \url{http://www.rc.ufl.edu} for providing computational resources and support that have contributed to the research results reported in this paper.

J.J.A.\ and E.K.\ acknowledge support for Program number (JWST-AR-04369.001-A) provided through a grant from the STScI under NASA contract NAS5-03127.

The POSYDON project is supported at Northwestern University by the Gordon and Betty Moore Foundation (grant awards GBMF8477 and GBMF12341), and at the University of Geneva by the Swiss National Science Foundation (grant awards CRSII5\_213497 and 2000-1-243011) and the Foundation for the University of Geneva (FUNIGE; grant award SWISS-ET).

This research was enabled by the computational resources and staff contributions provided by the Quest high-performance computing facility at Northwestern University, which is jointly supported by the Office of the Provost, the Office for Research, and Northwestern University Information Technology, hosting the Trident computing allocation (funded by the GBMF8477 and GBMF12341 grants) and the CIERA computing allocation (funded by CIERA). The computations were also performed at the University of Geneva using the Yggdrasil HPC services.
\end{acknowledgments}

\software{{\tt gaiamock} \citep{2024OJAp....7E.100E}, {\tt astropy} \citep{astropy}, {\tt NumPy} \citep{numpy}, {\tt matplotlib} \citep{matplotlib}}

\bibliographystyle{aasjournal}
\bibliography{gaia}
\end{document}